\documentclass[]{spie}  

\usepackage{amsmath,amsfonts,amssymb}
\usepackage{graphicx}
\usepackage[colorlinks=true, allcolors=blue]{hyperref}
\usepackage{aas_macros}

\title{DustRover: A Python Package for Modelling Dust Extinction Curves (Phase I)}

\author[1, 2]{Amir E. Bazkiaei}
\author[2, 3]{Tayyaba Zafar}
\author[1, 2]{Nuria P. F. Lorente}
\author[2, 3]{Anilkumar Mailvaganam}
\author[2, 3]{Arihant Raidani}
\affil[1]{Australian Astronomical Optics, Faculty of Science and Engineering, Macquarie University, NSW 2109, Australia}
\affil[2]{Astrophysics and Space Technologies Research Centre, Macquarie University, NSW, 2109, Australia}
\affil[3]{School of Mathematical and Physical Sciences, Macquarie University, NSW, 2109, Australia}

\authorinfo{Further author information: (Send correspondence to A.E.B.)\\A.E.B.: E-mail: amir.ebadati-bazkiaei@mq.edu.au}

\begin{document} 
\maketitle

\begin{abstract}
We introduce \texttt{DustRover}, an open-source Python package designed to model dust extinction curves using multi-wavelength regimes.
\texttt{DustRover} supports optical/infrared spectroscopy, broadband photometry, and X-ray data within a unified framework, enabling robust analysis of dust extinction curves in distant galaxies. 
The software features a modular architecture that strictly decouples data management, dust models, and Markov Chain Monte Carlo (MCMC) based statistical fitting. 
The package is highly user-oriented, using simple \texttt{YAML} configuration files and offering built-in visualisation tools for SEDs and extinction curves.
\end{abstract}

\keywords{\texttt{DustRover}, \texttt{MCMC}, Dust Extinction, Spectroscopy}
\section{INTRODUCTION}

We present phase I of \texttt{DustRover}, an open-source Python package developed to model dust extinction curves across diverse multi wavelength regimes~[\citenum{zafar_2025_17362935}]. 
Implementing validated,  peer-reviewed algorithms~[\citenum{2011A&A...532A.143Z, 2015A&A...584A.100Z}], \texttt{DustRover} simultaneously analyses diverse observational inputs.  
By integrating optical/infrared spectroscopy, broadband photometry, and X-ray data into a unified framework, \texttt{DustRover} enables robust analysis of dust extinction curves in distant galaxies.
The software features a modular architecture that strictly decouples data management, dust models, and Markov Chain Monte Carlo (MCMC)~[\citenum{2010CAMCS...5...65G}] based statistical fitting.
Designed with a strong emphasis on user accessibility, \texttt{DustRover} utilizes straightforward \texttt{YAML} configuration files and includes built-in visualisation tools for \texttt{SED}s and extinction curves.  

To bridge the gap between complex multi-instrument datasets and robust statistical modeling, the primary objective of \texttt{DustRover} is to provide a standardised, highly automated framework for extracting line-of-sight dust extinction curves.
Specifically, this framework is engineered to handle the unique challenges of modeling the environments of distant galaxies using \texttt{GRB}s and \texttt{QSO}s as background probes.
By seamlessly integrating disparate data formats, the software aims to eliminate the systematic bottlenecks traditionally associated with cross-instrument and multi-wavelength spectral energy distribution (\texttt{SED}) fitting..

Architecturally, the project focuses on transitioning away from monolithic, hard-coded scripts toward a modular, object-oriented ecosystem.
By migrating to Python, the package removes the need for expensive software licenses and leverages modern scientific computing libraries to provide an extensible platform suitable for modern, multi-instrument datasets
By enforcing a strict separation of concerns among data ingestion, dust models, and MCMC-driven statistical optimization, we aim to deliver an extensible interface where the community can introduce custom extinction parametrisations without refactoring the core codebase. Ultimately, our objective is to maximize workflow reproducibility through lightweight, human-readable \texttt{YAML} configurations and native visualisation suites, establishing a reliable pipeline for precise parameter estimation in dust analysis.

\section{SOFTWARE ARCHITECTURE AND IMPLEMENTATION}

\begin{figure} [htbp]
\begin{center}
\begin{tabular}{c}
\includegraphics[width=0.95\textwidth]{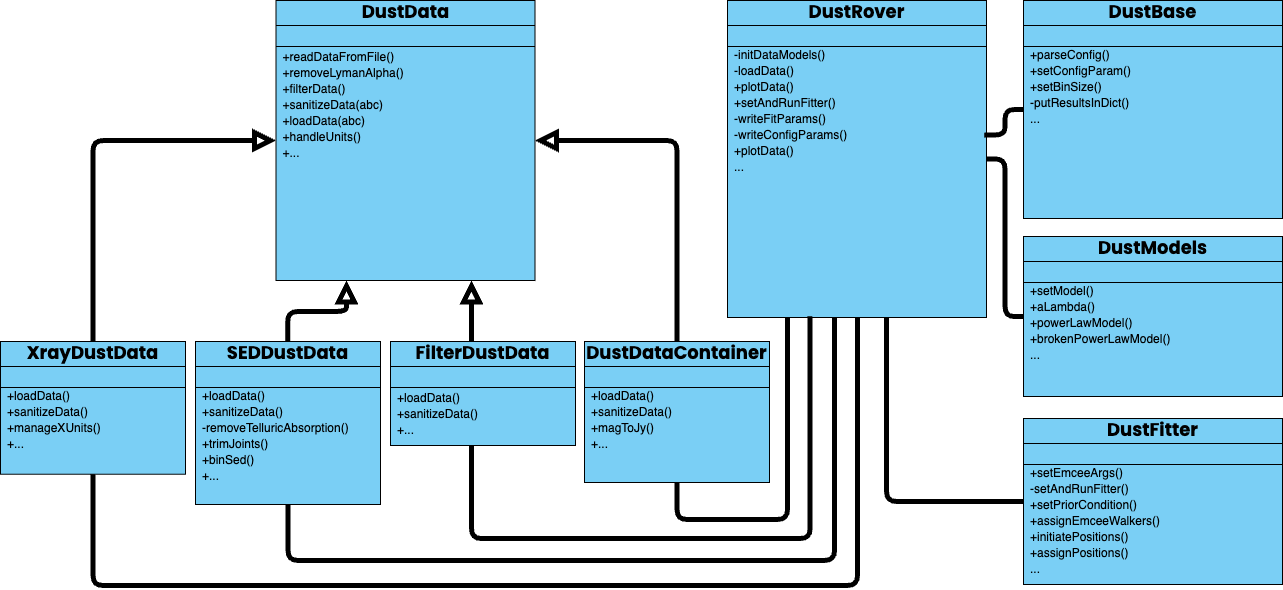} 
\end{tabular}
\end{center}
    \caption{\label{fig:uml}  \texttt{UML} class diagram showing the object-oriented architecture of \texttt{DustRover}. The main \texttt{DustRover} class acts as the central orchestrator, managing the pipeline execution workflow by coordinating three independent core modules: 1) a data management system based on \texttt{DustData}, using specialized subclasses (\texttt{SEDDustData}, \texttt{XrayDustData}, and \texttt{FilterDustData}) to ingest multi-wavelength inputs; 2) the \texttt{DustModels} module for physical extinction laws; and 3) the \texttt{DustFitter} engine, which handles the \texttt{MCMC} sampling via emcee~[\citenum{2013PASP..125..306F}] for parameter estimation and error propagation. Global configuration parsing is kept separate within the \texttt{DustBase} class. 
    }
\end{figure}

\texttt{DustRover} is designed as an object-oriented Python framework (Figure \ref{fig:uml}) that enforces a strict separation between data management, physical modeling, and statistical inference. The pipeline is driven by five core structural components coordinated by a top-level orchestrator.

\subsection{Orchestration (\texttt{DustRover})}
The top-level \texttt{DustRover} class acts as the central orchestrator of the entire ecosystem. 
It handles the underlying data objects, physical models, and fitting engines to manage the pipeline execution flow.
It sets up the execution by calling methods like \texttt{\_initDataModels()} and \texttt{loadData()}, runs the optimization via \texttt{setAndRunFitter()}, and finally triggers the built-in visualisation routines using \texttt{plotData()}.

\subsection{Configuration Base (\texttt{DustBase})}
To ensure reproducibility and clean environment management, the package derives its core configurations from the \texttt{DustBase} class. 
The \texttt{DustBase} class reads global settings, such as: object classification targets; numerical prior boundaries; data paths; and \texttt{MCMC} sampler steps, and converts them into a centralized data structure.
It initialises parameters via \texttt{parseConfig()} and tracks the active vs. fixed parameter space.
The user can run the entire pipeline using just a single \texttt{YAML} configuration file.
This file lets the user control all the high-level settings directly, so they never have to change or touch the underlying Python code.

\subsection{Data Management (\texttt{DustData})}

The data ingestion layer is built around the \texttt{DustData} abstract base class, which defines data cleaning protocols, handles coordinate units via \texttt{handleUnits()}, and cuts out contaminated physical regimes using routines like \texttt{removeLymanAlpha()}.
Three main data classes inherit from this base class and implement custom routines based on different dataset structures:

\begin{itemize}
    \item \textbf{\texttt{SEDDustData}:} Ingests optical and infrared spectroscopic inputs via \texttt{loadData()}, removes telluric absorption bands via \texttt{\_removeTelluricAbsorption()}, and automatically handles spectrograph arm alignment using \texttt{trimJoints()} and data down-sampling via \texttt{binSed()}.
    \item \textbf{\texttt{XrayDustData}:} Ingests high-energy spectroscopic observations, handling energy coordinate translations from $\text{keV}$ into wavelength space via \texttt{manageTxtUnits()}.
    \item \textbf{\texttt{FilterDustData}:} Manages broadband photometric inputs, translates observed magnitudes and corresponding errors into Janskys ($Jy$) using \texttt{magToJy()} flux conversion method.
\end{itemize}

\subsection{Extinction Modeling (\texttt{DustModels})}
The \texttt{DustModels} class isolates the analytical and physical descriptions of the implemented models from the rest of the pipeline.
Using the \texttt{setModel()} interface, the framework determines whether to apply standard transient power-law laws (\texttt{powerLawModel()}) or broken power-laws (\texttt{brokenPowerLawModel()}) for \texttt{GRBs}, or to construct a template-aligned background continuum for Quasars.
The module calculates the active cross-wavelength attenuation curve $A(\lambda)$ via \texttt{aLambda()}.
Because the extinction component is strictly decoupled from the data structures and fitting loops, users can implement custom extinction parametrisations by updating the core model interface without rewriting the underlying Python codebase.
This is done by simply updating the model parameters in the configuration \texttt{YAML} file.

\subsection{Statistical Fitting (\texttt{DustFitter})}
Statistical inference is completely decoupled from both the data structures and the physical models within the \texttt{DustFitter} module. 
This engine manages parameter estimation using an \texttt{MCMC} sampling framework integrated with the \texttt{emcee} package~[\citenum{2013PASP..125..306F}]. 
It translates active configuration boundaries into analytical boundaries using \texttt{setPriorCondition()} method, packages multi-instrument values via \texttt{setEmceeArgs()}, and initialises walker paths across the open parameter width space.

\section{PIPELINE VERIFICATION AND CONVERGENCE TESTS}

To verify the operational stability of the \texttt{DustRover} architecture, we execute a multi-wavelength validation run using a \textbf{GRB} dataset and a \texttt{QSO} dataset. In this work, we used observational dataset of the afterglow of \texttt{GRB 180325A} with a redshift of $z = 2.2486$ [\citenum{2018ApJ...860L..21Z, 2022MNRAS.517.6022C}] as an example of \textbf{GRB}. This data set includes multi-wavelength observations spanning from the X-ray to the infrared regimes. As a \texttt{QSO} example, we used the \texttt{SED} of \texttt{HAQ J0151+0618} with a redshift of $z = 0.95$ [\citenum{2015A&A...584A.100Z}].

The user initialises the \texttt{DustRover} class with the path to the configuration file as the input to the class. Then, they can run the \texttt{plotData()} method to display the cleaned data. 
That allows the user to visually inspect the data and, if required, revisit the cleaning configuration parameters such as \texttt{binSize} (for binning the \texttt{SED} data), \texttt{log10NVJoint} and \texttt{log10VUJoint} (for determining the joint frequency at which the corresponding \texttt{SED} data is cut). 

\begin{figure} [htbp]
\begin{center}
\begin{tabular}{cc}
\includegraphics[width=0.48\textwidth]{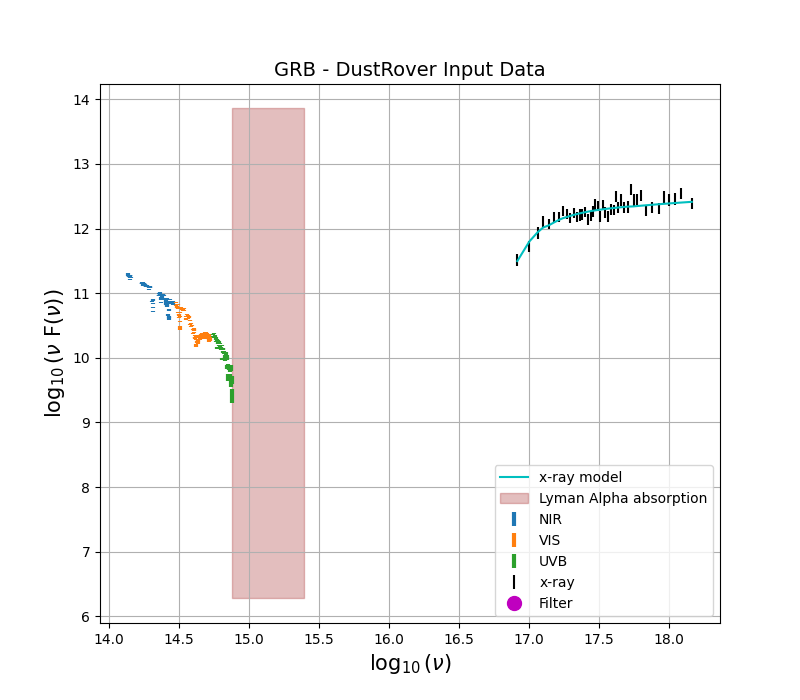} &
\includegraphics[width=0.48\textwidth]{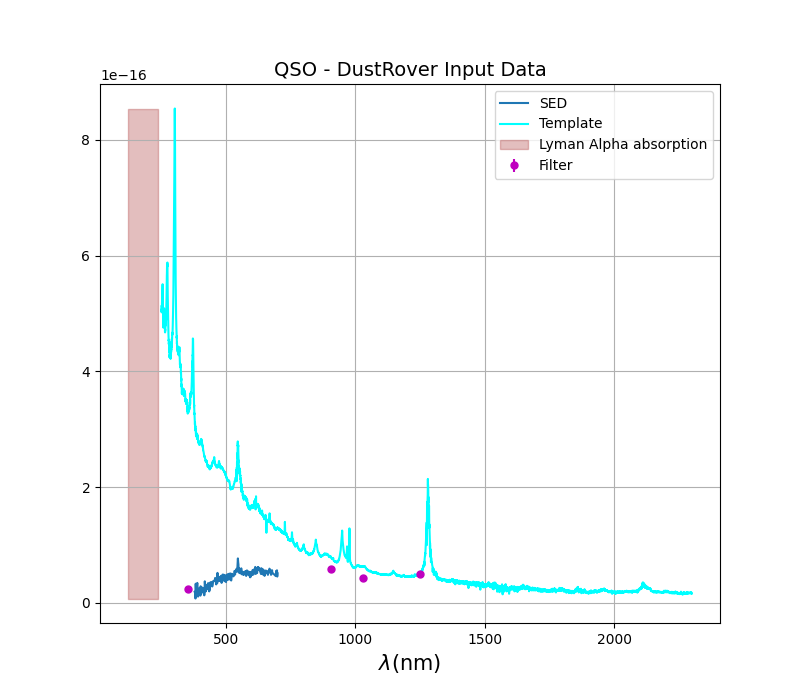}
\end{tabular}
\end{center}
\caption{\label{fig:cleaned_data} The data visualisation generated by the \texttt{plotData()} method displaying the ingested \texttt{GRB} (left panel) and \texttt{QSO} (right panel) datasets after preprocessing. The plot on the left overlays the independent, binned spectroscopic segments (\texttt{NIR}, \texttt{VIS}, and \texttt{UVB}), the designated Lyman-alpha absorption cutoff window (shaded region), and the high-energy X-ray data. The plot on the right shows the \texttt{SED} of the \texttt{QSO}, the scaled template, the photometry data, and the Lyman-alpha absorption cutoff window (shaded region).}
\end{figure}

As demonstrated in Figure~\ref{fig:cleaned_data}, the visualisation layer displays the dropped regions resulting from the automated atmospheric telluric cleaning and spectrograph joint trims (for \texttt{GRB}) as well as the broad Lyman-alpha absorption masking for both \texttt{GRB} and \texttt{QSO}.
By plotting the data, the visualisation module provides an immediate quality check, confirming that the automated data cleaning algorithms have properly preprocessed the diverse inputs before running the \texttt{MCMC} fitting chains.

\begin{figure} [htbp]
\begin{center}
\includegraphics[width=0.65\textwidth, height=0.50\textwidth]{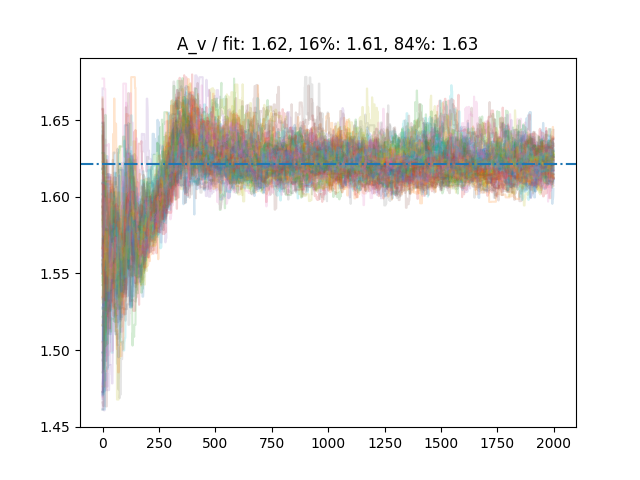}
\end{center}
\caption{\label{fig:mcmc_diagnostics} Trace plot tracking the execution paths of individual \texttt{emcee} walkers over 2,000 steps for parameter $A_v$. The dashed blue line marks the final best-fit value ($\mathbf{1.62^{+0.01}_{-0.01}}$) for the \texttt{GRB}.}
\end{figure}

After visual checks, the user starts the \texttt{MCMC} fitting chains by running \texttt{setAndRunFitter} method.
For the \texttt{GRB} run, the 2000 steps per walker was set via \texttt{mceeSteps} configuration parameter. 
Figure~\ref{fig:mcmc_diagnostics} displays the automated trace plot output for the visual extinction parameter $A_v$, demonstrating robust and rapid convergence to the posterior probability maximum.
The engine considers the last steps and extracts the median value along with asymmetric $1\sigma$ error margins from the 16th and 84th percentiles.
That is done using the last 100 steps in the chain by setting \texttt{chainDiscard} configuration parameter to 1900.
The resulting fit values ($A_v = 1.62$, $16\% = 1.61$, $84\% = 1.63$) are explicitly tracked and saved to the results.
These parameters are then mapped to reconstruct the final total-to-selective dust extinction curve shown in Figure~\ref{fig:mcmc_diagnostics} (Right), displaying a clean $1\sigma$ error boundary across the complete spectral range.
The resulting fit values ($A_v = 1.62$, $16\% = 1.61$, $84\% = 1.63$) are explicitly tracked and saved to the results.

Once the optimization loop completes, the orchestrator triggers the internal routines to automatically structure and store these outputs.
The framework creates a dedicated, timestamped directory including the best-fit values and percentiles stored in a \texttt{YAML} file. 
A second \texttt{YAML} file is created in the directory, which contains a copy of the original configuration settings to preserve full data provenance.
Additionally, the calculated attenuation values are written directly to a text table (\texttt{extCurve.txt}), providing a portable, publication-ready scientific record for downstream statistical combinations.

\begin{figure}[htbp]
\begin{center}
\includegraphics[width=0.85\textwidth]{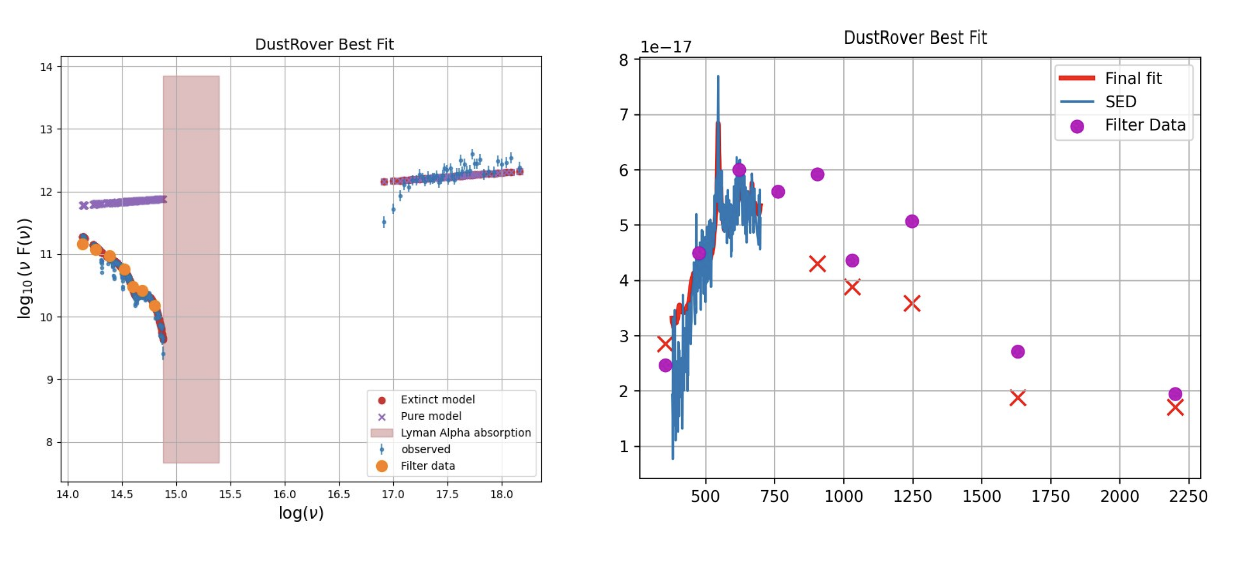}
\end{center}
\caption{\label{fig:grb_alignment} Left: Automated multi-wavelength data alignment plot showing the binned optical/infrared spectroscopic data (blue points on the left of the plot), discrete broadband photometric filter points (large orange markers), and high-energy X-ray observations (right-hand blue points data cluster) plotted alongside the best fit model (red points) for the \texttt{GRB}. Right: Spectral energy distribution (SED) and best-fit plot showing the spectroscopic data (blue line), broadband photometric filter observations (magenta circular markers) plotted alongside the final best-fit model spectrum (red thick line and crosses) for the \texttt{QSO}.}
\end{figure}

The final data alignment and model fit comparison is handled by the \texttt{plotFitResult()} method, displayed for both the \texttt{GRB} and the \texttt{QSO} in Figure~\ref{fig:grb_alignment}.
The left panel presents the observed \texttt{GRB} spectroscopic data, broad-band photometric data, and high-energy X-ray observations alongside the optimized models.
By overlaying the "Extinct model" (the attenuated background continuum in red) and the "Pure model" (the intrinsic, unreddened source continuum in purple), the visualization module lets the user directly inspect the determined dust parameters that account for the spectral attenuation across the optical and ultraviolet frequencies.
Similarly, the right panel demonstrates the method's output for the \texttt{QSO}, directly comparing the \texttt{SED} and broadband observations against the final best-fit model.
Ultimately, this automated visual output confirms the overall stability of the multi-instrument alignment and joint likelihood calculation across the \texttt{GRB} and the \texttt{QSO}.

\section{RESULTS AND FUTURE WORK}
The Phase I of \texttt{DustRover} successfully delivers a working, object-oriented Python pipeline for multi-wavelength dust extinction curve fitting.
The results are consistent when tested against \texttt{GRB} and \texttt{QSO} fits carried out via old methods.
Decoupling the data management, empirical dust models, and \texttt{MCMC} inference engine allow the user to produce parameter estimation and reproducible \texttt{SED} fits.  

Phase II will be focused on implementing a simultaneous grain model (silicates, graphite, \texttt{PAHs}, and iron) with variable depletion.
To handle this added physical complexity, we are re-engineering the backend with parallelized Bayesian algorithms and vectorized computation to drop runtimes from days to under an hour.
This high-performance optimization is critical for scaling the framework to handle the rapid-turnaround demands of modern large-scale transient programs.

\section{Acknowledgments}

We acknowledge the Traditional Custodians of the land on which Macquarie University is situated, the Wallamattagal people of the Dharug nation, and pay our respects to Elders past and present. 
This work was supported by MQ Research Acceleration Scheme 2023 (MQRAS) PURE ID 279036011.

\bibliography{report} 
\bibliographystyle{spiebib} 

\end{document}